\documentclass[12pt]{article}
\newcommand{\beao}{\begin{eqnarray*}}
\newcommand{\eeao}{\end{eqnarray*}}
\usepackage{a4}\usepackage{amsfonts}
\newcommand{\be}{\begin{equation}}
\newcommand{\ee}{\end{equation}}
\newcommand{\bea}{\begin{eqnarray}}
\newcommand{\eea}{\end{eqnarray}}
\newcommand{\beq}{\begin{eqnarray}}
\newcommand{\eeq}{\end{eqnarray}}
\newcommand{\nn}{\nonumber}
\newcommand{\pa}{\partial}
\newcommand{\ep}{\epsilon}

\newcommand{\Ref}[1]{(\ref{#1})}
\newcommand{\om}{\omega}\newcommand{\Om}{\Omega}
\renewcommand{\o}{\over}
\usepackage{graphicx}
\begin{document}
\title{The Casimir effect for thin plasma sheets and the role  of the surface plasmons}
\author{
{\sc M. Bordag}\thanks{e-mail: Michael.Bordag@itp.uni-leipzig.de} \\
\small  University of Leipzig, Institute for Theoretical Physics\\
\small  Vor dem Hospitaltore 1, 04103 Leipzig, Germany}
\maketitle
 \begin{abstract}We consider the Casimir force betweeen two dielectric bodies described by the plasma model and between two infinitely thin plasma sheets. In both cases in addition to the photon modes  surface plasmons are present in the spectrum of the electromagnetic field. 
We investigate the  contribution of both types of modes to the Casimir force and confirm resp. find in both models large compensations between the plasmon modes themselves and between them and the photon modes especially at large distances.
Our conclusion is that the separation of the vacuum energy into plasmon and photon contributions must be handled with care except for the case of small separations.
  \end{abstract}
 \section{Introduction}
 The Casimir force between macroscopic bodies is at present one of the most interesting macroscopic manifestations of   quantum effects. Applications range from attempts to obtain new constraints on hypothetical fifths forces in high precision experiments to impact on micromachinery and nanotechnology. The influence of the real structure of the interacting bodies is of increasing importance. For a better understanding of the Casimir forces efforts are undertaken to investigate   the contributions from separate parts of the spectrum of the electromagnetic field. For instance, it is assumed that that surface plasmons (waves propagating along the surface\footnote{The notion of surface plasmons is sometimes used in a different definition. So in \cite{Ger} it is generalized to include the photon modes too.}) dominate the Casimir force at small separations while at large separations the photon modes, i.e., the waves propagating perpendicular to the surfaces dominate. 

In the present paper we investigate the Casimir effect for thin plasma sheets and for  dielectric bodies described by the plasma model and can confirm the above assumption only partly. We find large cancelations between the plasmon modes and and between the plasmon and the photon modes at all distances. For the first model this was observed already in \cite{Lam}. \\
 
The considered models are:
\begin{enumerate}
\item {\it Two dielectric bodies described by the plasma model}\\
Actually we assume two half spaces filled with a dieelectric medium and separated by a plane gap of width $L$. The dielectric function is described by the plasma model,
\be\label{ep}\ep(\omega)=1-{\om_p^2\over \om^2},
\ee
where $\om_p$ is the plasma frequency.
\item {\it Thin plasma sheets}\\
Barton considered in \cite{BIII,BV}   a model of an infinitely thin shell filled with a charged fluid in order to describe a single  layer hexagonal structure  of carbon atoms aimed to model $C_{60}$ or carbon nano tubes. The interaction of such plasma sheet with the electromagnetic field results in a set of matching conditions for the latter. We consider two parallel sheets at separation $L$.
\end{enumerate}
In fact, these models result in very similar formulas.  Both can be described by matching conditions and the Casimir force is given by the Lifshitz formula or a simple modification of it. Both models have surface plasmons in their TM polarization. The first model is well known and nearly all quantities considered below can be found in literature, see e.g., \cite{Lam} or \cite{Bordag:2001qi} and references therein. We consider this model here in parallel to the second one which is less known in order to demonstrate the similarity between the two.

In general, the Casimir force can be represented as resulting  from the ground state energy of the electromagnetic field,
\be\label{E0}E_0=\sum_{(n)}{1\over 2}\hbar\om_{(n)},
\ee
where the sum runs over all quantum numbers, i.e., over the whole physical spectrum. In the considered cases it consists of the photon modes which propagate in the whole space. They are labeled by a wave vector $\vec{k}\in \mathbb{R}^3$ and a discrete index for the two polarizations, TE and TM. In addition for the  TM polarization there are the surface plasmons which are characterized by a discrete imaginary component of the wave vector in the perpendicular direction, $k_z=i\kappa$.
 
In general, internal degrees of freedom of the considered media, density oscillations in the second model for instance, give their own contributions to the vacuum energy. These contributions are not considered in the present paper.

Of course, $E_0$ as given by Eq. \Ref{E0} contains well known divergencies. These do not influence the Casimir force and can be removed for instance by subtracting the energy for infinite separation. After that \Ref{E0} is finite and represented by a converging sum/integral over photons and plasmons,
\be\label{div'n}E_0\equiv E_{\rm total}=E_{\rm plasmon}+E_{\rm photon}.
\ee
However, the integrand in $E_{\rm photon}$ is highly oscillating and very inconvenient to handle. Therefor it is natural to move the integration path to the imaginary axis where the integrand becomes exponentially decreasing and where it is easy to perform the integration  numerically. This is used in the Lifshitz formula too and gives directly $E_{\rm total}$.

Below we calculate the vacuum energy $E_{\rm total}$ in this way for both models. For the first model we repeat known results, for the second model the calculation is probably new. Then we calculate separately the plasmon contributions which is an easy exercise. From this the photon contribution is then $E_{\rm photon}=E_{\rm total}-E_{\rm plasmon}$ and we can compare the relative weight of the individual contributions. 
 
Technically our calculation is based on the formalism introduced in \cite{Bordag:1995jz} (see also in the review \cite{Bordag:2001qi}, section 5.1.1) where the vacuum energy is calculated in the background of a potential depending on one coordinate only. In this formulation the handling   of bound states is particularly  transparent. 

We start in the next section with a description of the models and a collection of the necessary formulas. In the third section we describe the formulas we use for the calculation of the vacuum energy. In the fourth section the calculations are carried out and the results are presented. Section 5 deals with the special case of two attractive delta potentials and   the last section contains the conclusions.

Starting from here we put $\hbar=1$ and consider $c$ as the speed of light relative to the vacuum.

\section{Description of the models}
Let us start from a scalar field $\Phi$ in a geometry of parallel planes. After Fourier transform in time and in the directions parallel to the planes we note the wave equation,
\be\label{weq}
\left({1\over c^2} \om^2-k_{||}^2+{\pa^2\over\pa z^2}\right)\Phi(k_{||},z)=0,
\ee
where $c$ is the speed of light which may depend on the medium. This equation is to be supplemented by the corresponding matching conditions at $z=\pm{L\o 2}$. 
In the following we need the scattering solutions to equation \Ref{weq} which are those solutions which have for $z\to\pm\infty$ the asymptotic behavior
\bea\label{asb}
 \Phi(k_{||},z) &\raisebox{-4pt}{${\sim\atop z\to - \infty}$} &\ e^{ikz}+r \ e^{-ikz}, \nn\\
&\raisebox{-4pt}{${\sim\atop z\to  \infty}$} &\ t \ e^{ikz},
 \eea
where $r$ and $t$ are  the reflection  and transmissision coefficients. The frequency of these solutions follows from the dispersion  relation $\om^2/c^2=k_{||}^2+k^2$.

A surface plasmon is a bound state in the one dimensional problem and appears on the positive imaginary axis at $k=i\kappa$ with a wave function after division by $t$
\bea\label{bs} \Phi_{\rm bs}(k_{||},z) &\raisebox{-4pt}{${\sim\atop z\to - \infty}$} &\ \frac{r}{t} \  e^{\kappa z}, \nn\\
&\raisebox{-4pt}{${\sim\atop z\to  \infty}$} &\  \ e^{-\kappa z} 
\eea
and $t$ has a pole in $k=i\kappa$ whereas ${r\o t}$ is finite. 
 
The matching conditions are as follows. 
\begin{enumerate}
\item {\it Dielectric bodies}\\
The field strengths components $D_\perp$ and $E_{||}$ must be continuous across the surface. The polarizations of the electromagnetic field can be separated into transverse electric (TE) and transverse magnetic (TM) ones. The matching conditions for the appearing scalar functions are 
\be\label{mcs1TE}\begin{array}{rcrlcrlr}
\Phi_+  &-&  \Phi_-&=&0, \qquad \Phi'_+\ - \ \Phi'_-&=&0, &\qquad(TE)\\
\ep_+ \Phi_+ &-&  \ep_- \Phi_-&=&0, \qquad \Phi'_+\ - \ \Phi'_-&=&0, &\qquad(TM)
 \end{array}\ee
where $\Phi_\pm$ and $\ep_\pm$ are the limiting values from the right and from the left.
The dispersion with $\ep(\om)$, Eq. \Ref{ep}, is
\be\label{disp1}\om=\sqrt{\om_p^2+k_{||}^2+k^2}.\ee
The reflection and scattering coefficients are well known. We note them first for one surface assuming $\ep\ne 1$ on the left side and $\ep=1$ on the right side.
 Because we have plane waves outside the surface we can write down the wave function directly,
\be\label{asc}
 \Phi(k_{||},z) =\left\{\begin{array}{lcl}
  \ e^{ikz}+r \ e^{-ikz} & \mbox{for}& z<0, \\
 \ t \ e^{iqz} &\mbox{for} & z>0 \end{array} \right.
 \ee
with $\om^2=k_{||}^2+q^2$. Then the coefficients are
\be\label{DTE1}\begin{array}{rcrlcrlr}
t&=&\frac{2k}{k+q}, & r=\frac{k-q}{k+q}, &\qquad(TE)\\ [8pt]
 t&=&\frac{2\ep k}{k+\ep q}, & r=\frac{k- \ep q}{k+\ep q} .&\qquad(TM)
 \end{array}\ee

\item {\it Plasma sheets}\\
For the plasma sheet the matching conditions demand $E_{||}$ to be continuous across the surface and for the normal component $\Delta E_z=\frac{\Om}{\om^2}\nabla_{||}E_{||}$ to hold, where $\Om$ is a parameter inverse proportional to the density of carriers in the plasma sheet. In this model the polarizations can be separated into TE and TM too and the matching conditions read then
\be\label{mcs2TE}\begin{array}{rcrlcrlr}
\Phi_+  &-&  \Phi_-&=&0, \qquad \Phi'_+\ - \ \Phi'_-&=&2\Om \ \Phi, &\qquad(TE)\\
\Phi'_+ &-&   \Phi'_-&=&0, \qquad \Phi_+\ - \ \Phi_-&=&-2\frac{\Om}{\om^2}\ \Phi'. &\qquad(TM)
 \end{array}\ee
The matching condition for the TE polarization is just the same as for a delta function potential on a plane with strength $\Om$. The potential is repulsive because $\Om>0$ follows from the physics of the model. As known, a repulsive delta potential does not accommodate bound states and so there are no surface plasmons in this model. However, below we will investigate the unphysical case of $\Om<0$ too where bound states are present. The matching condition for the TM polarization is sometimes associated with a $\delta$' potential although this is not correct (see for example \cite{Albeverio}, Appendix). The combination of parameters  $-2\frac{\Om}{\om^2}$ appears from the physics of the model and leads to a bound state resp. to a  surface plasmons. In this model the dispersion is simply
\be\label{disp2}\om=\sqrt{k_{||}^2+k^2}\ee
and the reflection and transmission coefficients are
\be\label{PTE1}\begin{array}{rcrlcrlr}
t&=&\frac{ik}{ik-\Om}, & r=\frac{\Om}{ik-\Om}, &\qquad(TE)\\ [8pt]
 t&=&\frac{\om^2}{\om^2+ik\Om}, & r=\frac{ik\Om}{\om^2+ik\om} .&\qquad(TM)
 \end{array}\ee
\end{enumerate}

For two parallel surfaces at distance $L$ the transmission coefficient written in terms of the coefficients from single surfaces is (a formula of this kind can be found e.g. in \cite{Reynaud})
\be\label{t2}t=\frac{t_lt_r\exp(i(q-k)L)}{1+\frac{t_l}{\overline{t}_l}\overline{r}_lr_r e^{2iqL}},
\ee
where $\overline{t}$ is the complex conjugated to $t$ and the indices indicate the left resp. right surface. This matters only for the first model where the coefficients \Ref{DTE1} are written for the left surface. That for the right one follow by interchanging $k$ and $q$ and substituting $\ep\to 1/\ep$.
In this way for the dielectric model
\be\label{DTE2}
\begin{array}{rcrlcrlr}
t&=&\frac{  \frac{4kq}{(k+q)^2}}{1-\left(\frac{k-q}{k+q}\right)^2 \ e^{2iqL}} &\qquad(TE)\\ [12pt]
 t&=&\frac{  \frac{4\ep kq}{(k+\ep q)^2}}{1-\left(\frac{k-\ep q}{k+\ep q}\right)^2 \ e^{2iqL}}  .&\qquad(TM)
 \end{array}\ee
holds and for the two plasma shells we have
\be\label{DPTE2}
\begin{array}{rcrlcrlr}
t&=&\frac{  \frac{-k^2}{(ik-\Om)^2}}{1-\left(\frac{\Om}{ik-\Om}\right)^2 \ e^{2ikL}} , &\qquad(TE)\\ [12pt]
 t&=&\frac{  \frac{\om^4}{(\om^2+ik\Om)^2}}{1+\left(\frac{\Om k}{\om^2+i\Om k }\right)^2 \ e^{2ikL}}  .&\qquad(TM)
 \end{array}\ee
The surface plasmons are given by the poles of $t$ on the imaginary axis at $k=i\kappa$. For the dielectric model we get the equation
\be\label{eqkaD}\frac{\kappa+\ep q}{\kappa- \ep q}=-\sigma e^{-qL}
\ee
with $\ep=1-\frac{\om_p^2}{\om_p^2+k_{||}^2-\kappa^2}$ and $q=\sqrt{k_{||}^2-\kappa^2}$. For $\sigma=+1$ the corresponding wave function is symmetric and for $\sigma=-1$ it is antisymmetric. The antisymmetric plasmon exists only for $k_{||}\ge \om_p/\sqrt{1+\om_p L/2}$. We denote the solutions of equation \Ref{eqkaD} by $\kappa_\sigma(k_{||},\om_p,L)$. We  introduce a similar notation for the frequency,
\be\label{fp}\om(k_{||},k)=\sqrt{\om_p^2+k_{||}^2+k^2}.\ee
A single surface with $\ep$ given by \Ref{ep} has one plasmon with
\be\label{kasi}\kappa_{\rm single}(k_{||},\om_p,L)=
\sqrt{\frac{\om_p^2}{2}+\sqrt{\left(\frac{\om_p^2}{2}\right)^2+k_{||}^4}},
\ee
which can be obtained from $\kappa_\sigma(k_{||},\om_p,L)$ in the limit $L\to\infty$.

For the plasma shell model we note the corresponding formulas. For the TE polarization the equation for   $\kappa$ is
\be\label{eqkaPTE}1+\frac{k}{\Om}=-\sigma e^{-kL}.   \ee
It has solutions for $\Om<0$ only and the antisymmetric solution has the additional constraint $|\Om|>1/L$. For one plane the solution is $\kappa_{\rm single}=-\Om$.
For the TM polarization the equation is
\be\label{eqkaPTEM} \frac{\om^2-\kappa \om}{\kappa\Om}=-\sigma e^{-\kappa L}  \ee
with $\om=\sqrt{k_{||}^2-\kappa^2}$. The solutions exist for $\Om>0$. For one plane the solution is $\kappa_{\rm single}=\frac{1}{2}\left(\sqrt{\Om^2+4k_{||}^2}-\Om\right)$.
Here we also introduce notations with an explicit indication of the arguments, $\kappa_\sigma(k_{||}, \Om,L)$ and 
\be\label{fp1} \om(k_{||}, k)=\sqrt{k_{||}^2+k^2}.\ee

Finally we note that for the plasmons the  inequalities
\bea\label{ineq}
\kappa_-<\kappa_{\rm single}<\kappa_+<\om_p&& \mbox{(dielectric plasma model),}\nn\\
\kappa_-<\kappa_{\rm single}<\kappa_+<k_{||}&& \mbox{(model of two plasma sheets)}
\eea
hold.

\section{The vacuum energy}
The vacuum energy for background potentials depending on one coordinate had been considered repeatedly. We follow here the derivation given in \cite{Bordag:1995jz}, or equivalently, in \cite{Bordag:2001qi}, section 5.1.1, but we change some notations, $s_{12}\to r$ and $s_{11}\to t$, for instance. The basic formula is
\be\label{E01} E_0=\frac12 \int\frac{dk_{||}}{(2\pi)^2} \
\left\{ \sum_\sigma \om(k_{||}, i\kappa_\sigma)+\int_0^\infty \frac{dk}{2\pi i} \ \om(k_{||},k) \frac{\pa}{\pa k}\ln \frac{t(k)}{t(-k)}
\right\},\ee
where $t(k)$ is the reflection coefficient defined in the scattering problem \Ref{asb} and the sum runs over the surface plasmons. This formula is quite general and a dependence of $\om$ and of $t(k)$ on $k_{||}$ is allowed. By means of $\frac{t(k)}{t(-k)}=e^{2i\delta}$ it is equivalent to representations in terms of scattering phase shifts or the mode density. 

In this formula the ultraviolet divergences are still present. We remove them by subtracting the doubled contribution from one single surface. This is equivalent to subtract the limit of large separation $L$. In the formulas for the transmission coefficients, Eqs. \Ref{DTE2} and \Ref{DPTE2}, this results in removing the factors in the numerators and in the plasmon contributions we substitute 
\be\label{subtr}  \om(k_{||}, i\kappa_\sigma)\to \om(k_{||}, i\kappa_\sigma)-\om(k_{||}, i\kappa_{\rm single}).
\ee
Obviously, we subtract distance independent terms which do not contribute to the Casimir force. 

By means of Eq. \Ref{E01} the vacuum energy is given as the sum \Ref{div'n} of plasmon and photon contributions and it consists of sum/integrals over the physical spectrum. While the (subtracted) plasmon contribution is fast converging, the (subtracted) photon contribution is highly oscillating. As already mentioned, it is sufficiently inconvenient to calculate it in this representation and we move the integration over $k$ towards the imaginary axis. In that case a fast converging integral appears which can be evaluated numerically without big effort. This procedure is standard and well known. For instance, the integration over the imaginary axis  is used in the  Lifshitz formula    although this is not always easy to see because of different choices of the  variables in the integrations. 

The only nontrivial moment which deserves attention is the role of the surface plasmons. In \cite{Bordag:1995jz} it was shown that in the process of deformation the integration contour crosses poles of the logarithmic derivative in \Ref{E01}. These are just the poles which correspond to the bound states resp. to the surface plasmons.  The crossings give extra contributions which just cancel the sum in \Ref{E01}. As a result one obtains the representation
\be\label{E02}E_0\equiv E_{\rm total}=\frac12 \int\frac{dk_{||}}{(2\pi)^2} \ \int_{k_0}^\infty \frac{dk}{\pi} \ 
\om(k_{||},ik) \frac{\pa}{\pa k}\ln  t(ik)
\ee
for the {\it total} vacuum energy. The lower integration boundary for $k$ is $k_0=\sqrt{\om_p^2+k_{||}^2}$ for the dielectric bodies with $\om(k_{||},ik)$ given by  \Ref{fp}  and $k_0=k_{||}$ for the plasma sheets with $\om(k_{||},ik)$ given by  \Ref{fp1}. This formula can be viewed also as a generalization of the Lifshitz formula to the case when   surface plasmons are present. 

Surface plasmons exist in both models in the TM polarization only. However,  in the TE polarization in the plasma shell model, which is equivalent to delta function potentials on the planes, surface plasmons exist for a 'wrong' sign of the charge density parameter $\Om$, i.e., for $\Om<0$. Although in the considered model this case is not physical it is interesting to consider it anyway. We do this in section \ref{sect5}.

\section{Calculation of the total vacuum energy and of the plasmon contributions}
As shown in the preceding section the total vacuum energy is given by Eq. \Ref{E02}.
For purposes of numerical evaluation it is meaningful to integrate by parts to get rid of the derivative. After that the formula for the total vacuum energy reads
\be\label{E03} E_{\rm total}=-\frac12 \int\frac{dk_{||}}{(2\pi)^2} \ \int_{k_0}^\infty \frac{dk}{\pi} \ 
\frac{k}{\om(k_{||},ik)} \  \ln  t(ik)
\ee
and this is the form which we use in the present section.

Now we collect the formulas for the transmission coefficients on the imaginary axis  taking the subtractions into account.
For the dielectric bodies
\bea \label{Dt}\left(t(ik)\right)^{-1}&=&1-\left({k-q \over k+q}\right)^2 \ e^{-2qL} ,\qquad (TE)
\nn \\
\left(t(ik)\right)^{-1}&=&1-\left({k-\ep q \over k+\ep q}\right)^2 \ e^{-2qL} ,\qquad (TM)
\eea
holds with $q=\sqrt{k^2-\om_p^2}$ and $\ep=1-\frac{\om_p^2}{\om_p^2+k_{||}^2-k^2}$
and for the plasma sheets we have
\bea \label{Pt}\left(t(ik)\right)^{-1}&=& 1-\left(\frac{\Om}{k+\Om}\right)^2 \ e^{-2kL} ,\qquad (TE)
\nn \\
\left(t(ik)\right)^{-1}&=&1-\left(\frac{\Om k}{k_{||}^2-k^2-\Om k }\right)^2 \ e^{-2kL} .\qquad (TM)
\eea
Inserting these formulas into \Ref{E03} the calculation of the vacuum energy is an easy numerical exercise. For $L\to\infty$ in all cases the ideal conductor limes $E=-\pi^2/240 \ L^{-3}$ is reobtained. 

Note that Eq.\Ref{E03} contains a double integral. In all cases one integration can be carried out explicitly.  In both models for the TE polarizations this is trivial because $t(ik)$ does not depend on $k_{||}$. For  the TM polarizations after changing the variables from $k_{||}$ to $\om$ the integration over $\om$ can be carried out explicitly.  This integration can be done by standard computer algebra programs and the result is a quite lengthy expression which we do not display here. Within the dielectric model the corresponding integration was done for the force, i.e., for the derivative of Eq.\Ref{E03} with respect to the distance, in the appendix in \cite{Lam}.

\begin{figure}\unitlength1cm
 \begin{picture}(14,5)
 \put(0,0){ \includegraphics[width=8cm]{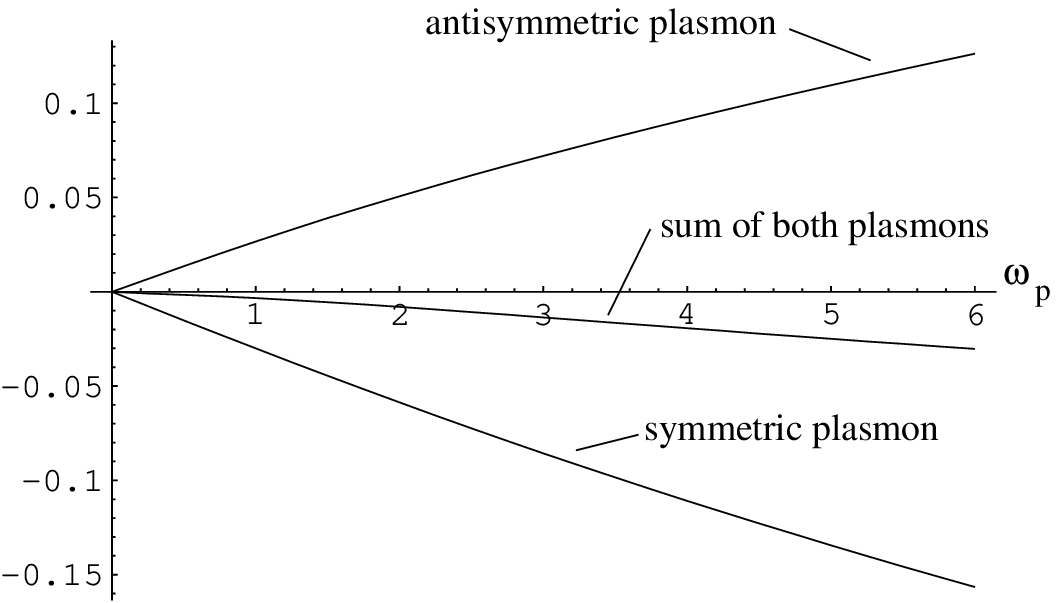}}
  \put(8,0){ \includegraphics[width=8cm]{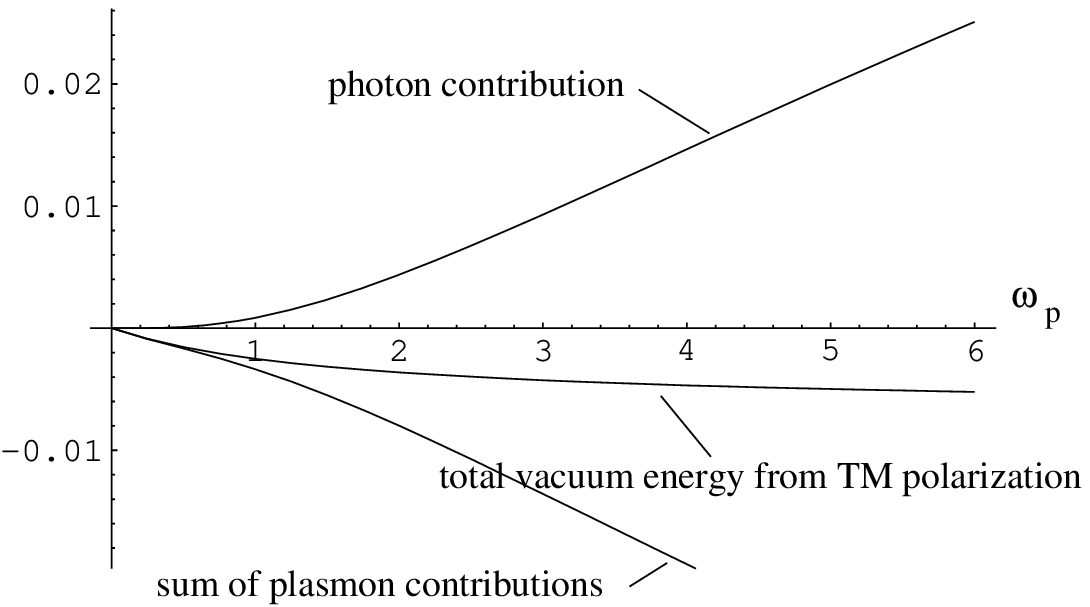}}\end{picture}
 \caption{The vacuum energy of the TM polarization in the dielectric plasma   model and its several constituents, observe the large compensation in both panels} \label{fig1}
\end{figure}

\begin{figure}\unitlength1cm
 \begin{picture}(14,4)
 \put(0,0){ \includegraphics[width=7cm]{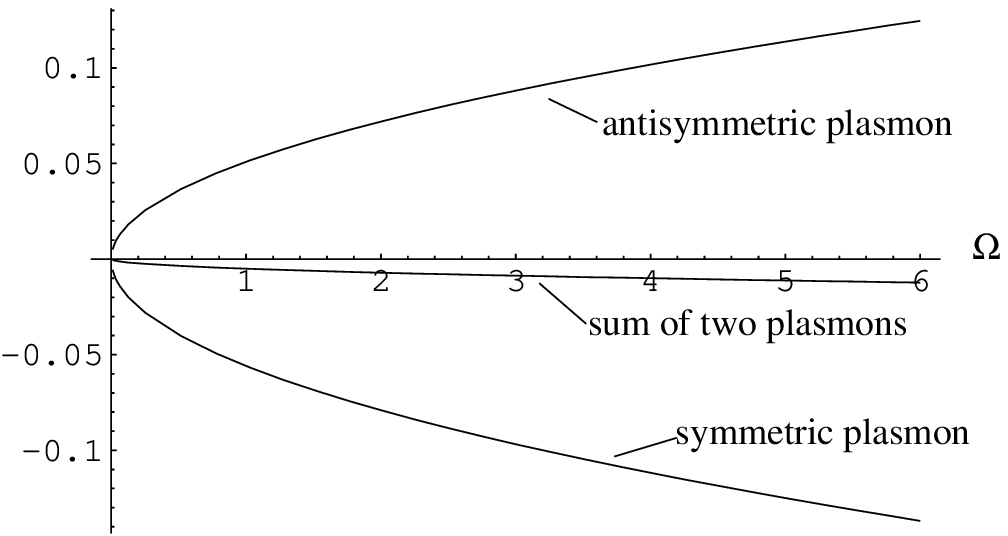}}
  \put(7,0){ \includegraphics[width=7cm]{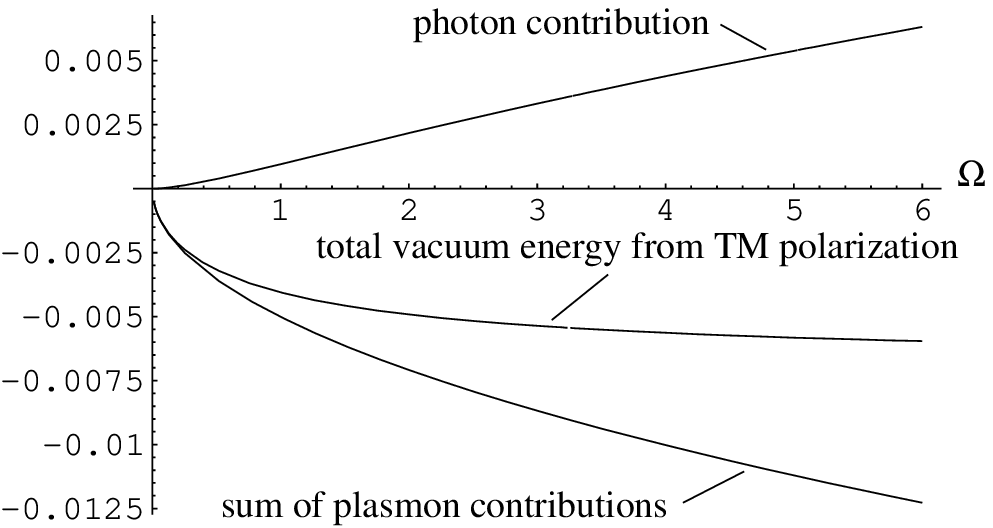}}\end{picture}
 \caption{The vacuum energy of the TM polarization in the plasma sheet model and its several constituents, observe the large compensation in both panels} \label{fig2}
\end{figure}

The contribution of the surface plasmons follows from Eq. \Ref{E01} and \Ref{subtr}. It reads
\bea\label{sd} E_{\rm plasmon}  &
=&\frac12 \int\frac{dk_{||}}{(2\pi)^2} \ \sum_{\sigma=\pm 1} \ 
\left\{\sqrt{\om_p^2+k_{||}^2-\kappa_\sigma^2(k_{||},\om_p,L)} \right.\nn \\ && \left. 
~~~~~~~~~~~~~ -\sqrt{\om_p^2+k_{||}^2-\kappa_{\rm single}^2(k_{||},\om_p,L)}   \right\}
\eea
for the dielectric bodies with the constraint $k_{||}\ge \om_p/\sqrt{1+\om_p L/2}$ for $\sigma=-1$  and 
\bea\label{sp} E_{\rm plasmon}  &
=&\frac12 \int\frac{dk_{||}}{(2\pi)^2} \ \sum_{\sigma=\pm 1} \ 
\left\{\sqrt{k_{||}^2-\kappa_\sigma^2(k_{||},\Om,L)} \right.\nn \\ && \left. 
~~~~~~~~~~~~~ -\sqrt{k_{||}^2-\kappa_{\rm single}^2(k_{||},\Om,L)}   \right\}
\eea
for the plasma sheets. All these quantities are shown in Fig. \ref{fig1} for the dielectric bodies as a function of the plasma frequency $\om_p$ and in Fig. \ref{fig2} for the plasma sheets as a function of the parameter $\Om$  for a fixed distance $L=1$. In both cases in the left panel we observe a significant compensation between the two plasmon contributions. In the right panel we see a compensation between the plasmon and the photon contributions.

\begin{figure}\unitlength1cm
 \begin{picture}(14,4)
 \put(0,0){ \includegraphics[width=7.3cm]{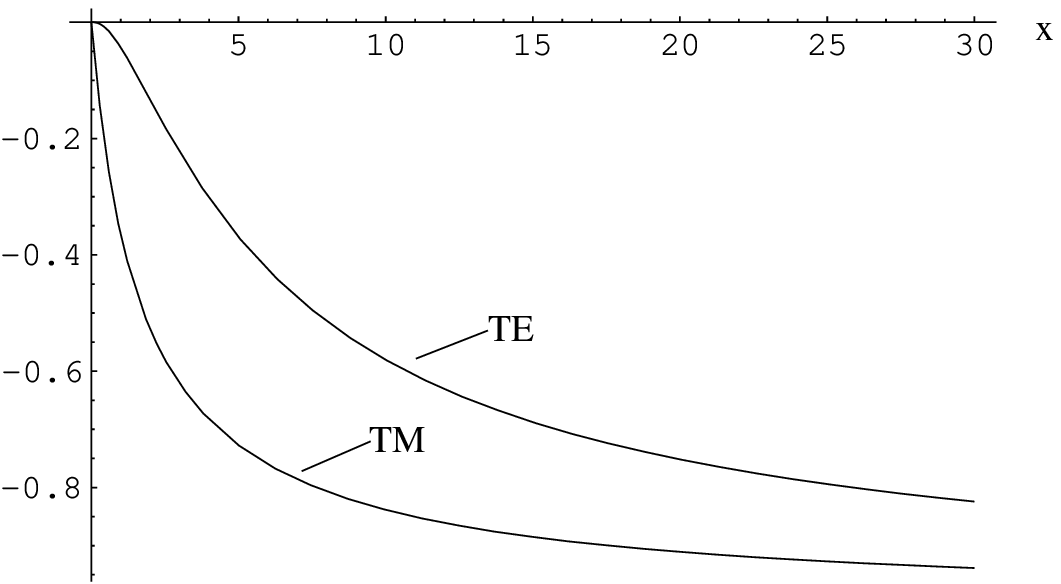}}
  \put(7,0){ \includegraphics[width=7cm]{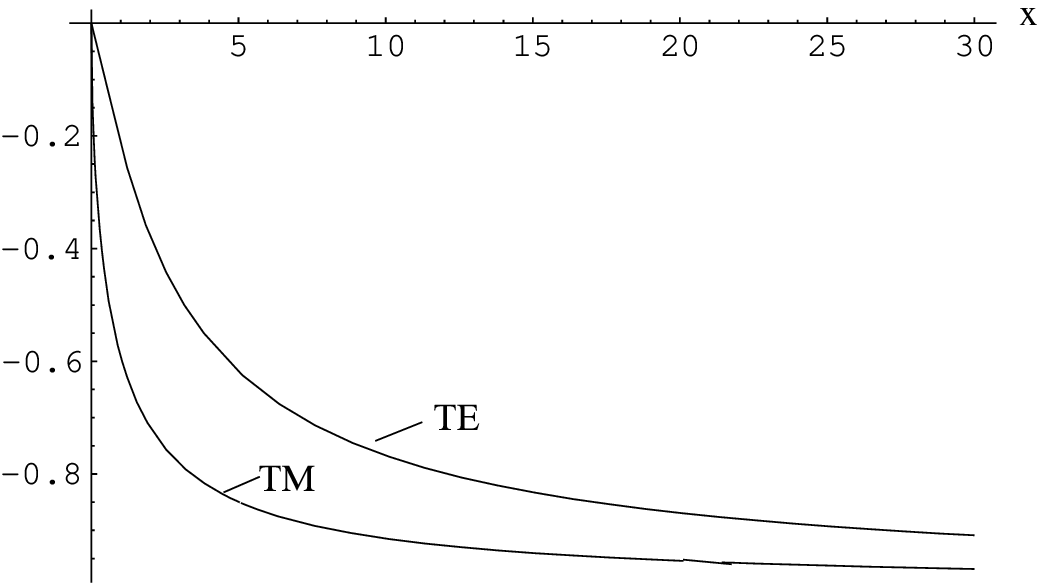}}
  \end{picture}
 \caption{The contributions from the TE and TM polarizations to the vacuum energy in the dielectric model (left panel) and in  the plasma sheet model (right panel) normalized to the perfect conductor limes} \label{fig4d1}
\end{figure}

The vacuum energy can be represented in the form
\be\label{dev} E_{\rm total}=-\frac{\pi^2}{240}\frac{1}{L^3} f(\om_pL),\ee
where the factor in front  is the ideal conductor contribution  and the dimensionless function $f(\om_pL)$ describes the deviation from the ideal conductor in the considered model. In case of the plasma sheet model it depends on $\Om L$. For large argument it has the limit
\be \lim_{x\to\infty}f(x)=1.\ee
For both models and both polarizations  this function is shown in Fig.\ref{fig4d1}. 

\begin{figure}\unitlength1cm
 \begin{picture}(14,4)
 \put(0,0){ \includegraphics[width=7cm]{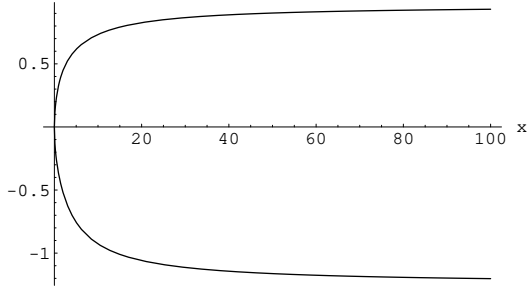}}
  \end{picture}
 \caption{The function $g(x)$ as defined in Eq.\Ref{g} in the dielectric plasma   model, the lower curve is for the symmetric plasmon and the upper curve is for the antisymmetric plasmon} \label{Abbg}
\end{figure}

Taking into account the dependence of the vacuum energy on the parameters as given by Eq.\Ref{dev}   we see from the figures Fig.\ref{fig1} and Fig.\ref{fig2} that the relative compensation between the plasmons and the photon contribution increases with distance L. This can be discussed in more detail. We consider the individual plasmon contributions as given by Eqs.\Ref{sd} and \Ref{sp} and represent these formulas in the form
\be\label{Epl}
E_{{\rm plasmon}, \ \sigma}=
\frac{1}{4\pi} \int_{k^0_{||}}^\infty dk_{||} \ k_{||} \ \ \om_\sigma(\om_p,L,k_{||})
\ee
with $k^0_{||}=\om_p/\sqrt{1+\om_pL/2}$ for $\sigma=-1$ in the dielectric model and $k^0_{||}=0$ otherwise. In \Ref{Epl} we introduced the notation
\be \om_\sigma(\om_p,L,k_{||})=\sqrt{\om_p^2+k_{||}^2-\kappa_\sigma(k_{||},\om_p,L)}  -\sqrt{\om_p^2+k_{||}^2-\kappa_{\rm single}(k_{||},\om_p,L)} .
\ee
The $\kappa_{\sigma}(k_{||},\om_p,L)$ are given by Eqs. \Ref{eqkaD} and \Ref{eqkaPTEM} and $\kappa_{\rm single}(k_{||},\om_p,L)$ by Eq.\Ref{kasi}. 
Now we make the substitution $k_{||}\to k_{||}/L$ in the integral in \Ref{Epl},
\be\label{Epl1}
E_{{\rm plasmon}, \ \sigma}=
\frac{1}{4\pi L^3} \int_{k^0_{||}L}^\infty dk_{||} \ k_{||} \ \ L\om_\sigma(\om_p,L,k_{||}/L).
\ee
For dimensional reasons the expression $L\om_\sigma(\om_p,L,k_{||}/L)$ depends on the product $\om_pL$ only. This is a representation in parallel to \Ref{dev} and the integral in \Ref{Epl1} is a dimensionless function on $\om_pL$. In the next step we substitute $k_{||}\to k_{||}\sqrt{\om_pL}$  and represent the plasmon contribution in the form
\be\label{Eplg}
E_{{\rm plasmon}, \ \sigma}=
\frac{\sqrt{\om_pL}}{4\pi L^3} \ g(\om_pL)
\ee
with 
\be\label{g}g(\om_pL)=\sqrt{\om_pL}\int_{k^0_{||}\sqrt{L/\om_p}}^\infty dk_{||} \ k_{||} \ \ L\om_\sigma(\om_p,L,k_{||}\sqrt{\om_p/L}).
\ee
So far this is a simple rewriting of the plasmon contribution. Now the   statement is that 
\begin{itemize}\item for the dielectric model the function $g(x)$ has a finite limit for $x\to\infty$, namely
\be\label{g1}\lim\limits_{x\to\infty}g(x)=\left\{
\begin{array}{rllll}
-1.2448 &  &\mbox{for}&  \sigma=1 &\mbox{(symmetric plasmon),}\\
0.9652 &   &\mbox{for} & \sigma=-1 &\mbox{(antisymmetric plasmon).}\end{array}\right.
  \ee
The function $g(x)$ is shown in Fig.\ref{Abbg} for both plasmons.
\item for the plasma sheets the function $g(x)$ does not depend on its argument at all. The values are
\be\label{g2}g(x)=\left\{
\begin{array}{rllll}
-0.702427&  &\mbox{for}&  \sigma=1 &\mbox{(symmetric plasmon),}\\
0.639449 &   &\mbox{for} & \sigma=-1 &\mbox{(antisymmetric plasmon).}\end{array}\right.
  \ee
  \end{itemize}
This is the  behavior of the plasmon contributions for large both or one of two, distance $L$ and plasma frequency $\om_p$ (in the second model $\om_p$ must be substituted by $\Om$). 
We see that the plasmon contributions grow by a factor $\sqrt{\om_pL}$ faster than the total vacuum energy. At once, by means of Eq.\Ref{div'n}, this gives us the asymptotic behavior of the photon contribution and we see how the mentioned compensation is growing.

It should be mentioned that in this way we got at once in the plasma sheet model the behavior of the plasmon contributions for small $\Om_p$ and for small $L$. 

We note that the statements of Eqs. \Ref{g1} and \Ref{g2}  are obtained numerically. No attempt had been made to get these results analytically. For the dielectric plasma model similar results are reported in \cite{Lam}. 

In a similar manner the asymptotic behavior  for small separations can be obtained numerically. In the dielectric model we find
\bea\label{Das}
E_{\rm TE}&\sim& -0.002 \ \om_p^3 ,\nn \\
E_{\rm TM}&\sim& -0.0039 \ \frac{\om_p}{L^2} , \nn \\
E_{\rm plasmon, \ \sigma=+1}&\sim& -0.0305 \ \frac{\om_p}{L^2}, \nn \\
E_{\rm plasmon, \ \sigma=+1}&\sim&  0.0267 \ \frac{\om_p}{L^2} . \eea
This confirms the known behavior  that in this limit  the TE contribution is small and the TM contribution is the sum of the two plasmon contributions. 

In the plasma sheet model we find
\beao
E_{\rm TE}&\sim& -0.013 \ \frac{\Om^2}{L} ,\nn \\
E_{\rm TM}&\sim& -0.00501 \ \frac{\sqrt{\Om}}{L^{5/2}} .\eeao
For the plasmons, by means of Eq. \Ref{g2}, we know the behavior for all distances,
\bea\label{42}
E_{\rm plasmon,\ \sigma=+1}&=&0.05090 \ \frac{\sqrt{\Om}}{L^{5/2}} ,\nn \\
E_{\rm plasmon, \ \sigma=-1}&=&-0.05589 \ \frac{\sqrt{\Om}}{L^{5/2}} .\eea
So in this model we observe the same pattern, the TE contribution is small and the TM contribution is at small separation the sum of the two plasmon contributions. 

\section{Two attractive delta potentials}\label{sect5}

As mentioned above for the wrong sign of $\Om$ in the plasma sheet model the TE polarization has surface plasmons too. This is equivalent to a scalar problem with attractive delta potentials on two planes. An single attractive delta potential has always one bound state. For two planes there are two of them, provided $L>\Om$ holds otherwise there is only one.  Further we note that, because the considered quantum field is massless, any bound state makes the potential instable, i.e., it causes  particle creation. As a consequence, the vacuum energy acquires an imaginary part. In general, there is nothing special about that and we include this case here only because we are discussing the role of the surface plasmons.

There are two ways to perform the calculations. The first is to make an analytic continuation in Eq.\Ref{E02}. When moving the binding energy $\kappa_\sigma$ of a plasmon which by means of \Ref{ineq} is initially below the integration part 
upwards so that $\kappa_\sigma$ becomes larger than $k_0$ than one needs to deform the integration path in order it not to hit the pole. Than the integral remains finite but it acquires an imaginary part. 

The second way is to calculate the plasmon and photon contributions separately. Than it is obvious that the photon contribution is real and the plasmon contribution delivers just the imaginary part. Indeed, consider a plasmon contribution,
\be\label{plal} E_{\rm plasmon}  
=\frac12 \int\frac{dk_{||}}{(2\pi)^2} \ 
\sqrt{k_{||}^2-\kappa^2}, \ee
where $\kappa$ is a solution of Eq.\Ref{eqkaPTE}.
As usual, we are concerned  with a divergent quantity. Here the subtraction \Ref{subtr} of the contributions of single planes does not work because $\kappa$ does not depend on $k_{||}$. So we take a standard regularization, say
\be\label{plreg} E_{\rm plasmon}^\delta  
=\frac12 \int\frac{dk_{||}}{(2\pi)^2} \ 
\sqrt{k_{||}^2-\kappa^2} \ e^{-\delta \sqrt{k_{||}^2-\kappa^2}}. \ee
This integral can easily be calculated explicitly,
\be\label{plreg1} E_{\rm plasmon}^\delta  
=\frac{1}{2\pi\delta^3}+\frac{i\kappa^3}{12\pi}+O\left(\delta\right). \ee
After dropping the distance independent divergent part we just get
\be\label{pl2} E_{\rm plasmon} 
= \frac{i\kappa^3}{12\pi}, \ee
which is the contribution of one surface plasmon for two attractive delta potentials.

\begin{figure}\unitlength1cm
 \begin{picture}(14,4)
 \put(0,0){ \includegraphics[width=7cm]{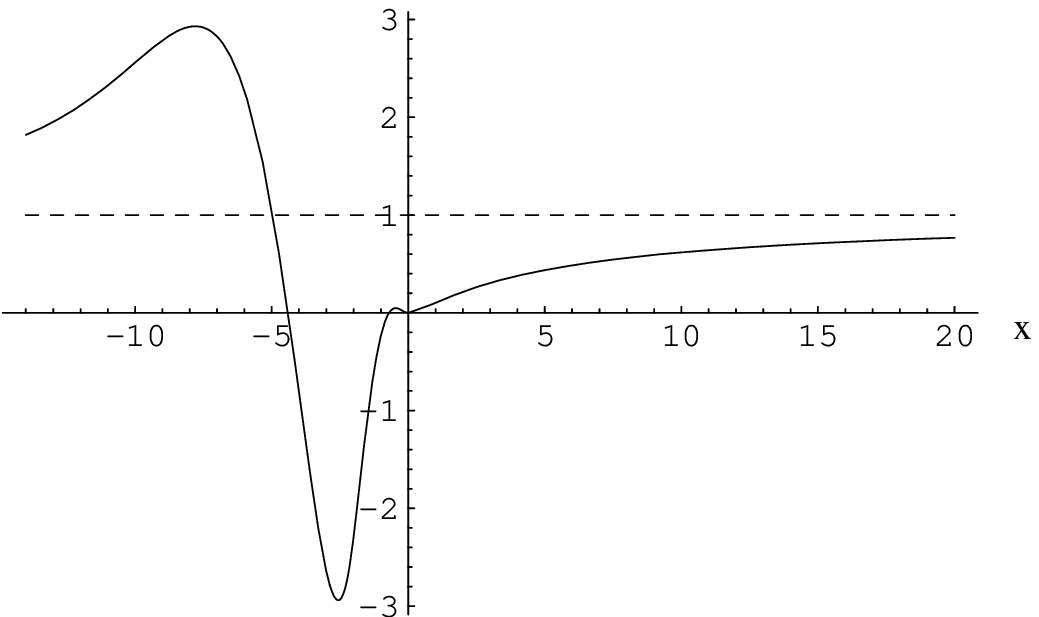}}
 \put(7,0){ \includegraphics[width=7cm]{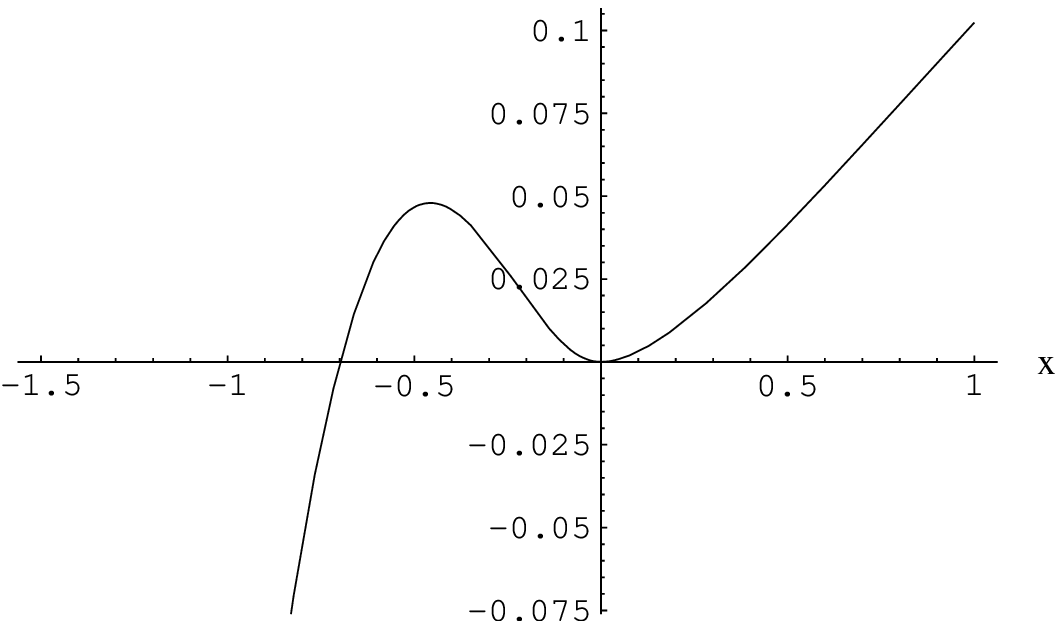}}
  \end{picture}
 \caption{In the left panel the function $h(x)$, \Ref{h}, which describes the vacuum energy relative to the ideal conductor limes for two delta potentials which are attractive for negative $x$ and repulsive for positive $x$ where they coincide with the TE polarization in the plasma sheet model, in the right panel the same function for small $x$ } \label{fig7}
\end{figure}

In the photon contribution we proceed in the same way as before. We move the integration path for $k$ and contract it on the imaginary axis. Now we have for $k_{||}<\kappa$ a pole on the cut. 
It can be shown, that after the contraction of the path, for example using the well known formula
$$ \frac{1}{x-i0}=\mbox{Vp}\frac{1}{x}+i\pi\delta(x),$$
the imaginary part disappears and only the Value principal integral remains which is real. Actually this happens because the factor $(k^2+k_{||}^2)^\frac{1}{2}$ has different signs on both sides of the cut. After that the order of integrations can be interchanged and the integration over $k_{||}$ can be carried out. Finally, integrating by parts, the photon contribution can be written in the form
\be\label{Eph2delta}E_{\rm photon}=\frac{-\pi^2}{1440 L^3} \ h(2\Om L)
\ee
with
\be\label{h}h(x)=\frac{45}{\pi^4}\ \mbox{Vp}\!\! \int_0^\infty dk \ k^2 \log \left| \frac{1}{1-\left(\frac{x}{k+x}\right)^2 \ e^{-k}} \right|.
\ee
The function $h(x)$ is shown in Fig.\ref{fig7}. It has the property $lim_{x\to\pm\infty}h(x)=1$. For $x>0$ it represents the deviation of the vacuum energy of the TE polarization in the plasma sheet model from the ideal conductor limit and for $x<0$ it represents the deviation of the real part of the vacuum energy of two attractive delta potentials.  We observe two regions with an attractive force and one with a repulsive force.

\section{Conclusions}
We calculated the vacuum energy in two models involving surface plasmons. As a result we are  able to support the physically instructive picture that at small distances the plasmon contributions dominate the vacuum energy.
However, there is a compensation by one order of magnitude between the two plasmon contributions (see Eq.\Ref{Das} and \Ref{42}). At large separation we find that the compensation between the plasmon contributions grows and that in addition there is a large and growing compensation between the sum of the two plasmon contributions and the photon contribution.

In the dielectric model the compensations were found in the paper \cite{Lam} where the importance of the plasmon contributions was emphasized. We would like to mention that 
the vacuum energy which is a relatively small quantity as compared to the plasmon contributions (even for  for small separations there is a compensation between the individual plasmons, see Eq.\Ref{Das})  can be directly calculated in the known representation where the momentum integration in the direction in perpendicular to the surfaces goes over the imaginary axis. In that formula large compensations are absent. A possible disadvantage, however, is that one needs to know $\ep(\om)$ for imaginary $\om$.

Also we considered the example of two planes carrying attractive delta potentials. Here the surface plasmons do not contribute to the real part of the vacuum energy. They deliver the imaginary part which is present because in that model particle creation is present. It is obvious that the imaginary part is present for large separations too. 

In general, the representations of the vacuum energy by an integral over the imaginary axis on the one hand side and as a sum of plasmon and photon contributions are connected by the dispersion relation the transmission coefficient $t(k)$ obeys, see for example Eq.(27) in \cite{Bordag:1995jz}. To some extend this is equivalent to the Kramers-Kronig relation for $\ep(\om)$. Both are exact relations and therefore a large compensation does not present a problem. However, if trying to make some kind of approximation in these relations the compensation may become a source of large errors.

\section*{Acknowledgements}
I thank V. Nesterenko and I. Pirozhenko for a number of interesting discussions. 
This work is supported by the grant NMP4-CT-2005-017071 within the EU STREP program.

\bibliographystyle{unsrt}\bibliography{../../Literatur/Bordag,../../Literatur/libri,PaperonPlasmoncontribution}
\end{document}